\documentclass[
 amsmath,amssymb,
 aps,prl,superscriptaddress,twocolumn,notitlepage,nofootinbib]{revtex4-1}

\usepackage{graphicx}
\usepackage[space]{grffile}
\usepackage{latexsym}
\usepackage{textcomp}
\usepackage{multirow,booktabs}
\usepackage{amsfonts,amsmath,amssymb}
\usepackage{url}
\usepackage{hyperref}
\hypersetup{colorlinks=false,pdfborder={0 0 0}}
\usepackage[utf8]{inputenc}
\usepackage[english]{babel}

\usepackage{graphicx,amsmath,amssymb,color,array}
\usepackage[normalem]{ulem}
\usepackage{multirow}
\usepackage{times}

\usepackage{amsfonts}
\usepackage{latexsym}
\usepackage{amsmath}
\usepackage{amssymb}
\usepackage{amsfonts}
\usepackage{amsthm}
\usepackage{mathrsfs}
\usepackage{natbib}
\usepackage{color,verbatim,graphics}
\DeclareMathAlphabet{\mathrsfs}{U}{rsfs}{m}{n}
\DeclareMathAlphabet{\mathpzc}{OT1}{pzc}{m}{it}
\DeclareMathAlphabet{\matheus}{U}{eus}{m}{n}
\DeclareMathAlphabet{\mathbbold}{U}{bbold}{m}{n}

\newcommand{\ba}{\begin{eqnarray}}
\newcommand{\be}{\begin{equation}}
\newcommand{\ee}{\end{equation}}
\newcommand{\beq}{\begin{equation}}
\newcommand{\eeq}{  \end{equation}}
\newcommand{\bea}{\begin{eqnarray}}
\newcommand{\eea}{  \end{eqnarray}}

\newcommand{\ea}{\end{eqnarray}}
\newcommand{\ban}{\begin{eqnarray*}}
\newcommand{\ean}{\end{eqnarray*}}

\newcommand{\tr}{\operatorname{tr}}

\newcommand{\ket}[1]{\left|#1\right\rangle}
\newcommand{\bra}[1]{\langle#1|}

\newcommand{\ketbra}[2]{|#1\rangle\langle#2|}

\newcommand{\ie}{{\it{i.e.}~}}

\newcommand{\rA}{\mathrm{A}}

\newcommand{\rB}{\mathrm{B}}

\newcommand{\tel}{\mathrm{tel}}

\newcommand{\cl}{\mathrm{cl}}

\begin{document}

\title{Experimental study of nonclassical teleportation beyond average fidelity}

\author{Gonzalo Carvacho}
\affiliation{Dipartimento di Fisica - Sapienza Universit\`{a} di Roma, P.le Aldo Moro 5, I-00185 Roma, Italy}

\author{Francesco Andreoli}
\affiliation{Dipartimento di Fisica - Sapienza Universit\`{a} di Roma, P.le Aldo Moro 5, I-00185 Roma, Italy}

\author{Luca Santodonato}
\affiliation{Dipartimento di Fisica - Sapienza Universit\`{a} di Roma, P.le Aldo Moro 5, I-00185 Roma, Italy}

\author{Marco Bentivegna}
\affiliation{Dipartimento di Fisica - Sapienza Universit\`{a} di Roma, P.le Aldo Moro 5, I-00185 Roma, Italy}

\author{Vincenzo D'Ambrosio}
\affiliation{ICFO-Institut de Ciencies Fotoniques, The Barcelona Institute of Science and Technology, 08860 Castelldefels (Barcelona), Spain}
\affiliation{Dipartimento di Fisica, Universit\`{a} di Napoli Federico II, Complesso Universitario di Monte S. Angelo, 80126 Napoli, Italy}

\author{Paul Skrzypczyk}
\affiliation{H. H. Wills Physics Laboratory, University of Bristol, Tyndall Avenue, Bristol, BS8 1TL, United Kingdom}

\author{Ivan \v{S}upi\'{c}}
\affiliation{ICFO-Institut de Ciencies Fotoniques, The Barcelona Institute of Science and Technology, 08860 Castelldefels (Barcelona), Spain}

\author{Daniel Cavalcanti}
\affiliation{ICFO-Institut de Ciencies Fotoniques, The Barcelona Institute of Science and Technology, 08860 Castelldefels (Barcelona), Spain}

\author{Fabio Sciarrino}
\email{fabio.sciarrino@uniroma1.it}
\affiliation{Dipartimento di Fisica - Sapienza Universit\`{a} di Roma, P.le Aldo Moro 5, I-00185 Roma, Italy}

\begin{abstract}
Quantum teleportation establishes a correspondence between an entangled state shared by two separate parties that can communicate classically and the presence of a quantum channel connecting the two parties. The standard benchmark for quantum teleportation, based on the average fidelity between the input and output states, indicates that some entangled states do not lead to channels which can be certified to be quantum. It was recently shown that if one considers a finer-tuned witness, then all entangled states can be certified to produce a non-classical teleportation channel. 
Here we experimentally demonstrate a complete characterization of a new family of such witnesses, of the type proposed in \textit{Phys. Rev. Lett. 119, 110501} (2017) under different conditions of noise. Furthermore, we show non-classical teleportation using quantum states that can not achieve average teleportation fidelity above the classical limit. Our results have fundamental implications in quantum information protocols and may also lead to new applications and quality certification of quantum technologies.
\end{abstract}

\maketitle

\textit{Introduction. --} The role of entanglement in quantum information processing is of utmost importance, but it is also subject of debate. 
Entanglement is today the core of many key discoveries ranging from quantum teleportation \cite{teleportation}, to quantum dense coding \cite{densecod}, quantum computation \cite{comp1,comp2,comp3} and quantum cryptography \cite{cryp1,cryp2}. Quantum communication protocols such as device-independent quantum key distribution \cite{DI1} are
 heavily based on entanglement to reach nonlocality-based communication security \cite{Bellreview}. 

The prototype for quantum information transfer using entanglement as a communication channel is the quantum teleportation protocol \cite{teleportation}, where a sender and a receiver share a maximally entangled state which they can use to perfectly transfer an unknown quantum state. This protocol represents a milestone in theoretical quantum information science \cite{NielsenChuang,gaussianqi,TelepReview} and lies at the basis of many technological application such as quantum communication via quantum repeaters \cite{quantumrepeater,quantumrepeater2} or gate teleportation \cite{quantumgate}. It has been implemented over hundreds of kilometers in free-space \cite{longdis1,longdis2} and more recently in a ground-to-satellite experiment \cite{groundtosatellite}. Employed platforms include mainly photonic qubits \cite{Bownmeester97,martini,lombardi,danubi,timebin1,timebin2,spin-orbit}, but also nuclear magnetic resonance \cite{NMR}, trapped atoms \cite{trapp1,trapp2}, atomic ensembles \cite{atomic1,atomic2} and solid-state systems \cite{solid1,solid2,solid3}.

To fully understand the role of entanglement in quantum teleportation, it is necessary to gauge what is the actual entanglement content (if any) that must be involved in order to upgrade a classical channel to a quantum channel.
Indeed, it is known that not all entangled states allow a teleportation fidelity to be reached which is higher than the one achievable using only classical communication \cite{Horodecki99}, with the notable example of bound entangled states \cite{Horoentangle,boundstat}. 
This mismatch between the nonclassicality of the shared state and nonclassicality of teleportation can be resolved by taking into account the full available information instead of only the teleportation fidelity. Very recently it has been theoretically shown \cite{PRLDA} that it is possible to use a more informative witness for teleportation scenarios which detects nonclassicality of a teleportation channel whenever any amount of entanglement is present in the shared state. 

\begin{figure}
\includegraphics[width=1\columnwidth]{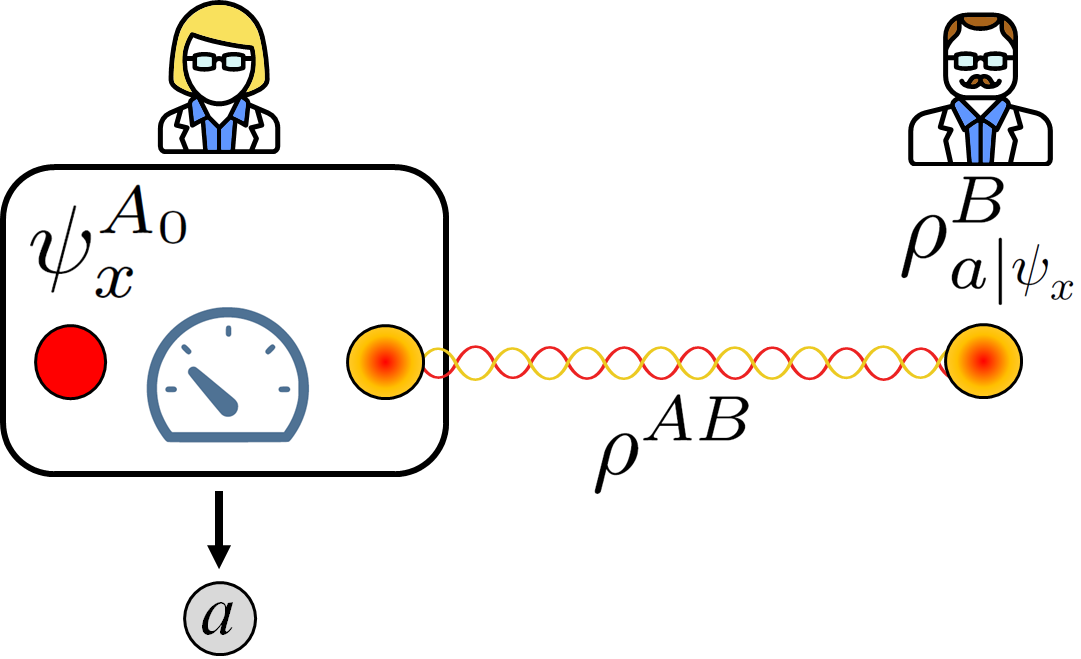}
\caption{Teleportation scenario: Alice and Bob use a shared system in a quantum state $\rho^{\rA\rB}$. Alice measures her subsystem together with system $A_0$ that can be prepared in states $\ket{\psi_x^{A_0}}$. Conditioned on Alice's measurement outcome $a$ Bob's system is left in the state $\rho^B_{a|\psi_x}$.}
\label{fig:teleportation}
\end{figure}

\begin{figure*}[t]
\includegraphics[width=\textwidth]{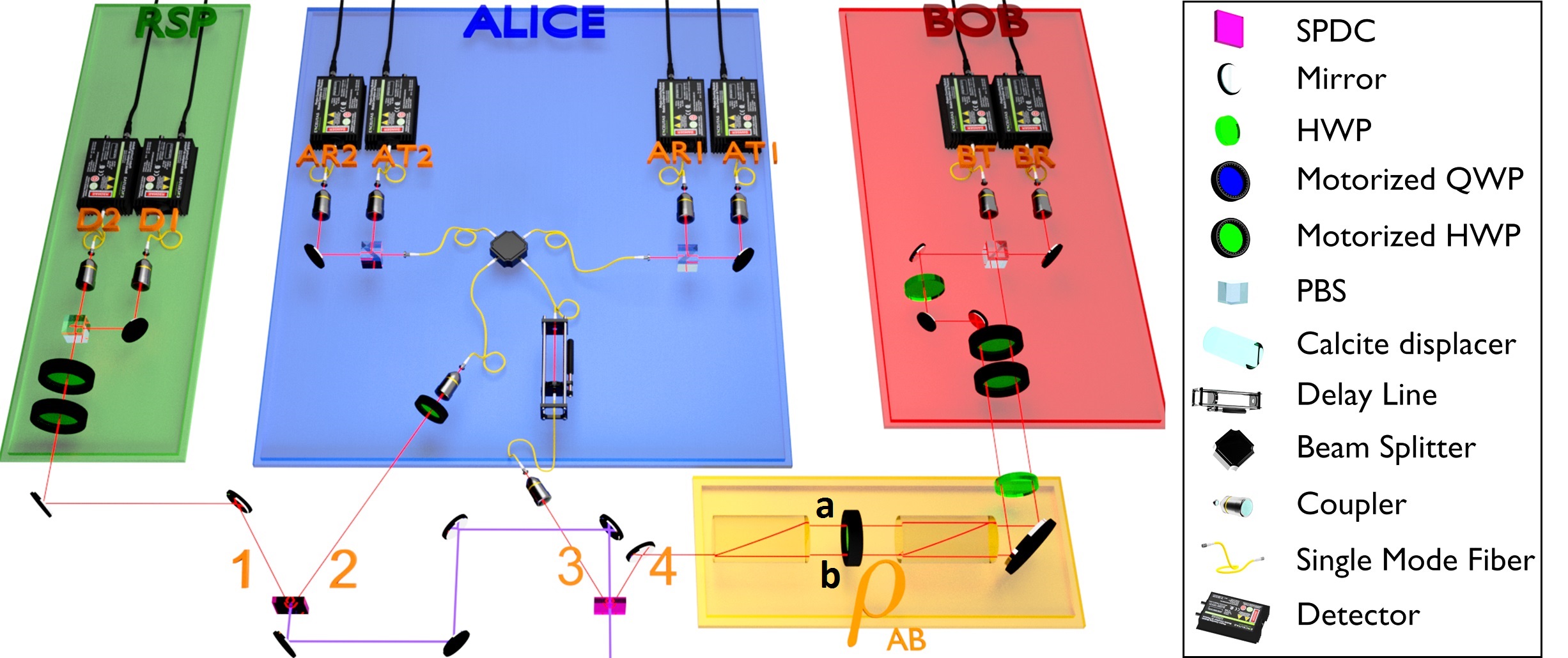}
\caption{{\bf Experimental apparatus}: Two photon pairs (1-2) and (3-4) are generated via parametric down-conversion in two separated nonlinear crystals. Remote state preparation (RSP) of photon 2 is performed by projective measurement on photon 1 via quarter wave-plate, half wave-plate and polarizing beam splitter. Photon 4 is sent through an interferometer composed by a first calcite displacer (CD), a half wave-plate (HWP) with a tunable angle $\alpha$, a second CD and a HWP at $45^o$. When $\alpha$ = $45^o$ photon 4 comes out in path 4a without modifications of the joint state $\rho^{\rA\rB}$. Conversely, when $\alpha = 0^o$, photon 4 comes out in path 4b projected onto the vertical polarization $\ket{V}$, and the joint state $\rho^{\rA\rB}$ becomes $\ket{VV}$. Intermediate angles between $0^o$ and $45^o$ allow to tune the two-qubit state $\rho^{\rA\rB}$ between $\ket{\psi^-}$ and $\ket{VV}$. The two paths 4a and 4b are incoherently recombined at the PBS of Bob's measurement stage by means of a HWP at $45^o$ on path 1b. To test nonclassicality of teleportation depending on the amount of noise, Alice implements a Bell state measurement on photons 2 and 3 by sending them into a 50/50 beam splitter (BS) and performing polarization analysis of the two outputs. A motorized delay line (DL) ensures temporal indistinguishability in the BS.}
\label{fig:Exp}
\end{figure*}

In this article, we experimentally test, using pairs of polarization-entangled photons, the properties of a non-classical teleportation witness by exploiting a photonic teleportation setup able to generate a family of channel states (shared between the sender and the receiver) characterized by a tunable amount of noise $\gamma$. Moreover, we study the behavior of the non-classical teleportation witness for various values of $\gamma$ and we check its agreement with theoretical predictions. By considering a particular model for the additional non-tunable experimental noises which affect any photonic teleportation setup similar to ours, we also address the correspondence between the witness results and the expected presence of entanglement in the channel state. Finally, we experimentally address how the non-classical teleportation witness is able to demonstrate the presence of a quantum channel even in conditions where the allowed maximum teleportation fidelity is below the classical limit. Our experimental results show that the analyzed non-classical teleportation witness represents a novel tool which is able to certify the presence of a non-classical teleportation channel beyond the possibilities of the previously adopted benchmark.

\textit{Non-classical teleportation witness.-- }In a teleportation protocol two parties, Alice and Bob, share a quantum state $\rho^{\rA\rB}$, which they want to use to teleport arbitrary (possibly unknown) states $\ket{\psi_x^{{\rA_0}}}$ (see Fig. \ref{fig:teleportation}). Alice applies a measurement $M$ with measurement operators $M_a^{\rA_0 \rA}$ on the systems $\rA_0 \rA$ in her possession, leaving Bob's system in the states
\be\label{assemblage}
\rho_{a|\psi_x}^\rB=\frac{\tr_{\rA_0\rA} [(M_{a}^{\rA_0\rA}\otimes \openone^\rB)\cdot(\ketbra{\psi_x}{\psi_x}^{\rA_0}\otimes\rho^{\rA\rB})]}{p(a|\psi_x)},
\ee
where $p(a|\psi_x)=\tr [(M_{a}^{\rA_0\rA}\otimes \openone^\rB)\cdot(\ketbra{\psi_x}{\psi_x}^{\rA_0}\otimes\rho^{\rA\rB})]$ is the probability of the particular outcome $a$. 

As proposed in \cite{PRLDA}, a particular witness function which exploits the full information obtained from teleportation can be computed in order to certify the presence of entanglement in the channel state $\rho^{\rA\rB}$. In particular, we focus here on the situation where the channel state belongs to the family of quantum states 
\begin{equation}
\rho^{\rA\rB}(\gamma)=\gamma\ket{\psi^-}\bra{\psi^-} + (1-\gamma)\ket{11}\bra{11},
\label{sharedstate}
\end{equation}
where $\ket{\psi^-}=(\ket{01}-\ket{10})/\sqrt{2}$, and as input to the teleportation experiment we consider the six eigenvectors of the Pauli matrices $\{\psi_x^{A_0}\}_{x=0}^5=\{ (\ket{0}+\ket{1})/\sqrt{2}, (\ket{0}-\ket{1})/\sqrt{2}, (\ket{0}+i\ket{1})/\sqrt{2}, (\ket{0}-i\ket{1})/\sqrt{2}, \ket{0}, \ket{1} \}$.  We denote the resulting set of states prepared for Bob $\rho_{a|\psi_x}^\rB(\gamma)$. 

\begin{figure*}
\includegraphics[width=\textwidth]{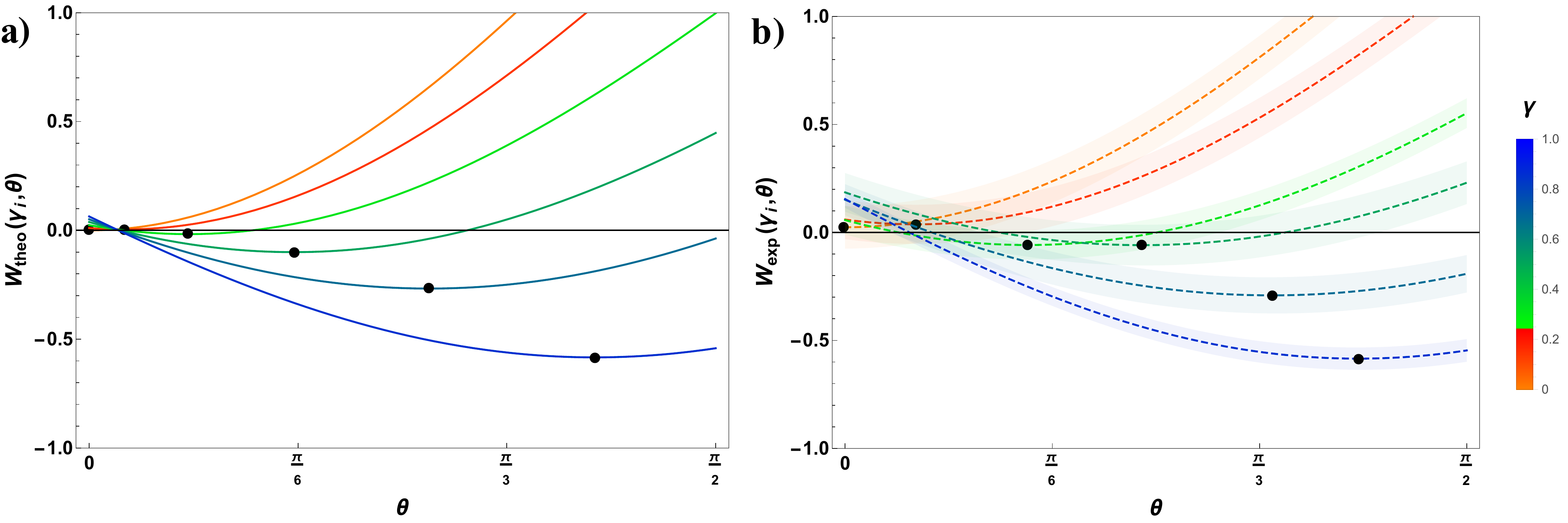}
\caption{{\bf Experimental test of the non-classical teleportation witness}. 
Theoretical (lines in figure 3-a) and experimental (dashed lines of figure 3-b) witness function obtained assuming our noise parameter estimation and varying the free parameter $\theta$, given different values of $\gamma$. Shaded areas correspond to one standard deviation of uncertainty upper and lower the dashed experimental lines, due to poissonian statistics. The bar legend on the right explains the color relationship with $\gamma$, while black points on the curves stands for the minimum experimental value obtained for a given $\gamma$. }
\label{fig:Data1}
\end{figure*}

Using the methods discussed in Ref. \cite{PRLDA}, one can obtain operators $F_{a|\psi_x}^\rB(\theta)$, defining a one-parameter family of non-classical teleportation witnesses: 
\begin{equation}\label{witness family}
W(\gamma,\theta)=\sum_{a,x} p(a|\psi_x) \tr[F_{a|\psi_x}^\rB(\theta)\, \rho_{a|\psi_x}^\rB(\gamma)],
\end{equation}
where the operators $F_{a|\psi_x}^\rB(\theta)$ are given in Table \ref{t:werner states} and the parameter $\theta$ identifies a single witness of the family.
It turns out that $\sum_{a,x} p(a|\psi_x) \tr[F_{a|\psi_x}^\rB(\theta)\, \rho_{a|\psi_x}^\rB]\geq0$ for all sets $\{\rho_{a|\psi_x}^\rB\}$ that come from teleportation processes using only classical communication (or separable states). Thus, $W(\gamma,\theta)<0$ is a certificate of nonclassical teleportation.

Moreover, we compare this witness with the average fidelity of teleportation, which is the commonly adopted as benchmark for the quality of quantum teleportation. The average fidelity between the input and output states of the process \cite{TelepReview}, is given by
$\overline{F}_{\tel}=1/|x|\sum_{a,x}p(a|\psi_x)\bra{\psi_x} U_a \rho^B_{a|\psi_x} U_a^{\dagger} \ket{\psi_x},$
where $U_a$ are fixed (\ie, independent of the input states) unitary operators conditioned to Alice's result $a$. In the case of perfect teleportation (using a maximally entangled state), $\overline{F}_{\tel}=1$, while in experimental situations it is always the case that $\overline{F}_{\tel}<1$. If one defines $\overline{F}_{\cl}$ as the maximal fidelity that can be obtained given a classical channel (i.e. in the absence of entanglement between Alice and Bob), the observation of $\overline{F}_{\tel}> \overline{F}_{\cl}$ implies that the teleportation process has no classical counterpart \cite{MassarPopescu95}. On the other hand, some entangled states cannot lead to a $\overline{F}_{\tel}$ higher than the classical bound. 
This can be the case for states in the family \eqref{sharedstate}, which lead to an average fidelity lower than the classical bound of $2/3$ when $\gamma\leq 1/2$ (see Supplementary Material), while the channel state is still entangled. Notably, for these states, the capability to achieve nonclassical teleportation can be still certified through the witness $W(\gamma,\theta)$.

\begin{table}[b]
$\psi_x$ \vspace{0.1cm}\\
$a$\hspace{0.1cm}\,
\begin{tabular}{c||c|c|c|c}
	$F_{a|\psi_x}^\rB$ & $\frac{\ket{0} + \ket{1}}{\sqrt{2}}$ & $\frac{\ket{0} - \ket{1}}{\sqrt{2}}$ & $\frac{\ket{0} + i\ket{1}}{\sqrt{2}}$ & $\frac{\ket{0} - i\ket{1}}{\sqrt{2}}$  \\ \hline\hline
0 & $-2\sin \theta \sigma_x$ & $2\sin \theta \sigma_x$ & $-2\sin \theta \sigma_y$ & $2\sin \theta \sigma_y$  \\ \hline
1 & 0 & 0 & 0 & 0 
\end{tabular}

$\psi_x$ \vspace{0.1cm}\\
$a$\hspace{0.1cm}\,
\begin{tabular}{c||c|c}
	$F_{a|\psi_x}^\rB$ & $\ket{0}$ & $\ket{1}$  \\ \hline\hline
0 & $4(1 -\cos \theta)\ket{1}\bra{1}$ & $4(1 +\cos \theta)(\ket{0}\bra{0})$ \\ \hline
1 & 0 & 0
\end{tabular}
\caption{\label{t:werner states} One-parameter family of teleportation witnesses operators ($\theta \in (0,\pi/2]$) used in eq. \eqref{witness family} that detects the nonclassicality of the teleportation process using the state in eq. \eqref{sharedstate}, the input states $\psi_x$ and a partial BSM where the outcome $a=0$ corresponds to a projection into a singlet state $\ket{\psi^-}=(\ket{01}-\ket{10})/\sqrt{2}$, and $a=1$ the orthogonal subspace.}
\end{table}

In the following, we experimentally study the behavior of $W(\gamma,\theta)$, characterizing it with respect to the experimental imperfections arising in a photonic setup, and then we  finally compare its application with the the average fidelity of teleportation benchmark.

\begin{figure*}[t]
\includegraphics[width=\textwidth]{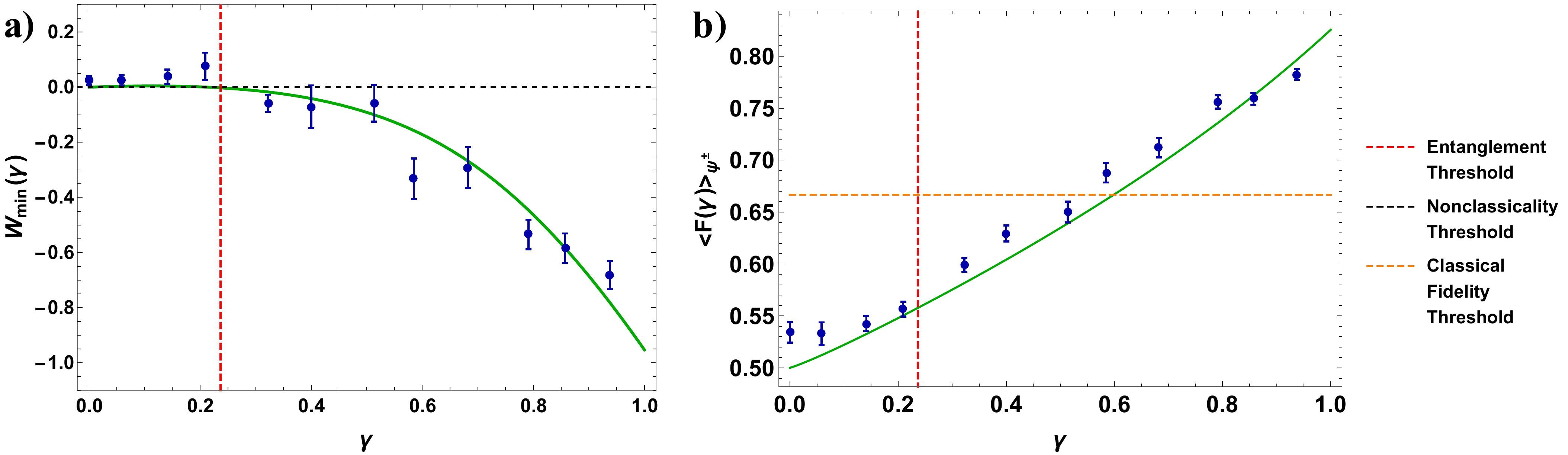}
\caption{{\bf Experimental application of non-classical teleportation witness} a) Optimized witness as a function of the channel parameter $\gamma$, corresponding to the dots from Fig. 3-b. Green line represents the theoretical prediction of $W_\mathrm{min}(\gamma)$, accordingly to our noise model (see Supplemental Material), while blue points show the optimal experimental values of the non-classical teleportation witness. The black dashed line is the non-classical teleportation threshold (i.e. the maximal value of $W_\mathrm{min}(\gamma)$ necessary to certify nonclassical teleportation), while the red dashed line represents the minimal value of $\gamma$ needed to have an entangled teleportation channel, given our noise model. 
b) Average fidelity of teleportation estimation. The green line shows the theoretical estimation for the mean value curve $\langle F(\gamma)\rangle_{\psi^{\pm}}$ (given our noise model), while blue points depict the experimental values. The red dashed line represents the minimum value of $\gamma$ necessary to have entanglement in the teleportation channel, while the orange one shows the bound for the classical fidelity of teleportation.
Error bars indicate one standard deviation of uncertainty, due to poissonian statistics.}
\label{fig:Data2}
\end{figure*}

\textit{Characterization of the non-classical teleportation witness.-- }We experimentally tested the proposed non-classical teleportation witness of eq. \eqref{witness family} by implementing a teleportation scenario in which a tunable amount of noise can be introduced in the entangled pair shared between Alice and Bob (see Fig. \ref{fig:Exp}). Two photon pairs (1-2 and 3-4) are generated in two separated nonlinear crystals by means of type-II spontaneous parametric down-conversion (SPDC) process. Photon pair 3-4 will embody the quantum channel $\rho^{\rA\rB}$ shared by Alice and Bob. To generate the tunable amount of noise we send Bob's photon through the interferometer depicted in the yellow rectangle of Fig. \ref{fig:Exp}, which is composed by a first calcite displacer (CD), a half wave-plate (HWP) at an angle $\alpha$, a second CD and a HWP at $45^o$. When $\alpha=45^o$ the interferometer implements the identity transformation on Bob's photon, while for $\alpha=0^o$ the state is vertically projected $\ket{V}$. As a result, in the ideal case Alice and Bob will share a state of the form of \eqref{sharedstate}, where the parameter $\gamma$ can be tuned by acting on the physical angle $\alpha$. The relationship between $\alpha$ and $\gamma$ is explicitly derived in the Supplementary Material. This ideal situation, however, cannot be completely reproduced in any actual implementation. As discussed in the Supplementary Material, to account for this we develop a model in order to theoretically predict the main imperfections which can arise in a photonic scheme similar to the one in Fig. \ref{fig:Exp}. We address the SPDC generation and the calcite interferometer as the two main sources of imperfection, and we introduce two parameters ($0\leq v \leq 1$ and $0\leq \delta\leq 1$, respectively) in order to take them into account. Given our model, we estimate that the state produced in the experiment, to good approximation, has the form
\begin{widetext}
\begin{equation}
\label{eq:noisy_state}
\rho_{\text{noisy}}^{\rA\rB}=
\dfrac{1}{4} \begin{pmatrix}
(1-v)\gamma & 0 &0 & 0\\
0 & 2 -v(2-3\gamma)-\gamma & -2(1-2\delta)^2 v \gamma & 0\\
0& -2(1-2\delta)^2 v \gamma & (1+v)\gamma &0\\
0&0&0& 2+v(2-3\gamma)-\gamma
\end{pmatrix}.
\end{equation}
\end{widetext}
As a result, given the estimated values of $v$ and $ \delta$, for some low values of $\gamma$ the generated quantum states can become separable, differently from the ideal one shown in eq. \eqref{sharedstate}.

We were able to experimentally test the expected properties of the nonclassicality witness of eq. \eqref{witness family}, analyzing its behavior for different values of $\gamma$ and $\theta$. 
In our scheme, Alice prepares the state to be teleported in photon 2 by means of remote state preparation ({\it i.e.} by performing projective measurement on photon 1) and teleporting it to Bob by performing a partial Bell state measurement (BSM) on photons 2 and 3. BSM is implemented with an in-fiber $50/50$ beam splitter (BS) followed by polarization analysis on each of the two outputs. Bell states $\ket{\psi^+}$ and $\ket{\psi^-}$ will be identified by two-fold coincidences from the same output arm and from different output arms, respectively. We address the ``robustness'' of the witness by considering the six eigenstates of the Pauli matrices $\{\psi_x^{A_0}\}_{x=0}^5$ as the possible states to be teleported. We measured different basis combinations of the teleported states $\rho^\rB_{a|\psi_x}$ and used this data to experimentally estimate the value of $W(\gamma,\theta)$. \\

Experimental results are shown in Fig. \ref{fig:Data1}-b, where the estimated value of $W(\gamma,\theta)$ is plotted as a function of the parameter $\theta$ and for different levels of noise $\gamma$. Fig. 3-a shows the theoretical expectations for the same set of states and parameters for comparison. As expected the witness certifies the non-classicality of the channel even in those cases when teleportation fidelity is lower than classical fidelity. This is shown in detail in figure 4 a-b.

In Fig. 4-a are reported the optimal values of W for fixed amounts of $\gamma$. Blue points show the experimental results, obtained evaluating $W(\gamma,\theta)$ for different $\gamma$-s and numerically minimizing over the parameter $\theta$. As expected, only those points with $\gamma$ higher than the estimated entanglement threshold (red dot-dashed line), have $W_{min}(\gamma)<0$, and thus show a nonclassical teleportation behavior.
The data are in good agreement with the theoretical values of $W_{min}(\gamma)$ (green line) which are expected given our imperfections model (see Supplementary Material).
To have a direct comparison with the usual teleportation witness we plot in Fig. 4-b the estimated teleportation fidelity for the same set of experimental points.

\textit{Discussion.}-- Our work experimentally demonstrates the complete characterization of a recently introduced non-classical teleportation witness \cite{PRLDA}. We started with the generation of a general family of two-qubit states unable to display a $\overline{F}_{\tel}$ within the quantum regime. We tuned the amount of noise in our states in order to test the properties of the nonclassical witness for different values of $\gamma$ and $\theta$ and establishing in this way a direct and general link between the expected properties and our experimental results. Moreover, we experimentally showed that beyond a certain noise threshold one can enter a region where the standard benchmark of the average fidelity is useless to certify a quantum channel while our witness can, thus providing an experimental 
tool with fundamental implications in quantum information protocols which can also lead to new applications in quantum technologies.

\textit{Acknowledgements.}--This work was supported by the ERC-Starting Grant 3D-QUEST (3D-Quantum Integrated Optical Simulation; grant agreement number 307783): http://www.3dquest.eu, ERC CoG QITBOX, Ram\'on y Cajal fellowship (Spain), Spanish MINECO (QIBEQI FIS2016-80773-P and Severo Ochoa SEV-2015-0522), Fundacio Cellex, Generalitat de Catalunya (SGR875 and CERCA Program) and the Royal Society (URF UHQT). G.C. thanks Becas Chile and Conicyt for a doctoral fellowship.

\end{document}


\title{Supplemental Material: Experimental test of nonclassical teleportation beyond average fidelity}

\author{Gonzalo Carvacho}
\affiliation{Dipartimento di Fisica - Sapienza Universit\`{a} di Roma, P.le Aldo Moro 5, I-00185 Roma, Italy}

\author{Francesco Andreoli}
\affiliation{Dipartimento di Fisica - Sapienza Universit\`{a} di Roma, P.le Aldo Moro 5, I-00185 Roma, Italy}

\author{Luca Santodonato}
\affiliation{Dipartimento di Fisica - Sapienza Universit\`{a} di Roma, P.le Aldo Moro 5, I-00185 Roma, Italy}

\author{Marco Bentivegna}
\affiliation{Dipartimento di Fisica - Sapienza Universit\`{a} di Roma, P.le Aldo Moro 5, I-00185 Roma, Italy}

\author{Vincenzo D'Ambrosio}
\affiliation{ICFO-Institut de Ciencies Fotoniques, The Barcelona Institute of Science and Technology, 08860 Castelldefels (Barcelona), Spain}

\author{Paul Skrzypczyk}
\affiliation{H. H. Wills Physics Laboratory, University of Bristol, Tyndall Avenue, Bristol, BS8 1TL, United Kingdom}

\author{Ivan \v{S}upi\'{c}}
\affiliation{ICFO-Institut de Ciencies Fotoniques, The Barcelona Institute of Science and Technology, 08860 Castelldefels (Barcelona), Spain}

\author{Daniel Cavalcanti}
\affiliation{ICFO-Institut de Ciencies Fotoniques, The Barcelona Institute of Science and Technology, 08860 Castelldefels (Barcelona), Spain}

\author{Fabio Sciarrino}
\affiliation{Dipartimento di Fisica - Sapienza Universit\`{a} di Roma, P.le Aldo Moro 5, I-00185 Roma, Italy}

\maketitle

\section{Methods}

\subsection{State Generation Protocol}
The core of the teleportation channel shared between Alice and Bob is constituted by the following quantum state

\begin{equation}
\rho_{AB}=\gamma\ket{\psi^-}\bra{\psi^-} + (1-\gamma)\ket{VV}\bra{VV}
\label{sharedstate}
\end{equation}

This peculiar quantum state has been experimentally obtained by pumping one of two entangled photons into a particular interferometric apparatus, which exploits the birefringence properties of calcites.\\
Let us start by considering the singlet state generated via a SPDC process. In order to properly take into account the effect of a calcite on one of the two photons we can introduce an additional path degree of freedom ($\tau_j$) for that photon. We thus obtain that the quantum state $\ket{\Phi_{AB}}$ after the SPDC generation and before the first calcite is given by

\begin{equation}
\label{eq:state_before_calcite}
\ket{\Phi_{AB}}=\dfrac{1}{\sqrt{2}}(\ket{H}\ket{V,\tau_1}-\ket{V}\ket{H,\tau_1}).
\end{equation}

A calcite will act on a photon coupling the polarization and the path degree of freedom, through the operator

\begin{equation}
\label{eq:calcite}
\mathcal{O}_{calc.}=\mathbb{I}\otimes(
\displaystyle \sum_{i=1}^{\infty}
\ket{H,\tau_{i}}\bra{H,\tau_{i}}+
\ket{V,\tau_{i+1}}\bra{V,\tau_{i}}),
\end{equation}

where we have to take into account that when a photon assumes a path value $\tau_j$ for $j\geq 3$, then it exits the interferometer and it gets lost. The whole interferometer is composed of a sequence of a calcite, a half-wave plate at angle $\alpha$, another calcite and a final half-wave plate at angle $\pi/4$, all acting on the proper photon of \eqref{eq:state_before_calcite}. The overall effect will then be given by:

\begin{equation}
\mathcal{O}_{interf.} = U_{\lambda/2}(\dfrac{\pi}{4}) \;\mathcal{O}_{calc.} \;U_{\lambda/2}(\alpha)\; \mathcal{O}_{calc.},
\end{equation}

where we defined the action of the half-wave plate on the proper photon as

\begin{equation}
U_{\lambda/2}(\alpha)=\mathbb{I}\otimes \left(\begin{array}{cc} \cos 2\alpha & \sin 2 \alpha \\ \sin 2 \alpha & - \cos2 \alpha \end{array}\right)\otimes\mathbb{I}.
\end{equation}

This process leads to the preparation of the final quantum state:
\begin{equation}
\begin{array}{c}
\rho_{AB}= \Tr_{\tau_i}[\mathcal{O}_{interf.}\ket{\Phi_{AB}}\bra{\Phi_{AB}}\mathcal{O}^{\dagger}_{interf.}]=\\
\\
\gamma\ket{\psi^-}\bra{\psi^-} + (1-\gamma)\ket{VV}\bra{VV},
\end{array}
\end{equation}

where
\begin{equation}
\label{eq:gamma_def}
\gamma=\dfrac{4 \sin^2 2\alpha}{3-\cos 4 \alpha}.
\end{equation}

\subsection{Noise Model}
We studied the behaviour of the nonclassicality witness with respect to the experimental noise arising in our setup. 
We considered, in our model, two different sources of noise. \\
First of all we took into account the SPDC generation of the two entangled couples of photons, and we dealt with this point by considering the presence of white noise. Since our goal is to analyse the behaviour of the witness with respect to the presence of entanglement within the teleportation channel, we addressed all the white noise to the channel state, including the noise owed to the remote state preparation of the teleporting states. The state generated immediately after the SPDC crystal can be thus assumed to be:
\begin{equation}
\rho^{SPDC}_{AB}= (v^2 \ket{\psi^-}\bra{\psi^-}+(1-v^2)\dfrac{\mathbb{I}}{4}),
\end{equation}

instead of a pure singlet state.\\
The second source of noise that we considered belongs to the manipulation of the photons in the calcites interferometer, necessary to obtain the state of Eq. \ref{sharedstate}.

 We took into account that each calcite can be affected by a dephasing process, and thus we elaborated a noisy model by considering the passage through a calcite as an open system evolution given by:
\begin{equation}
\mathcal{O}_{calc.}^{noisy}[\rho]=\Delta[\mathcal{O}_{calc.}\rho \mathcal{O}^{\dagger}_{calc.}],
\end{equation}

where $\mathcal{O}_{calc.}$ is defined in 
\begin{equation}
\label{eq:calcite}
\mathcal{O}_{calc.}=\mathbb{I}\otimes(
\displaystyle \sum_{i=1}^{\infty}
\ket{H,\tau_{i}}\bra{H,\tau_{i}}+
\ket{V,\tau_{i+1}}\bra{V,\tau_{i}}),
\end{equation}

the dephasing is given by:
\begin{equation}
\Delta[\rho]=\delta \rho + (1-\delta)\; \mathbb{I}\otimes\sigma_z\otimes\mathbb{I}\; \rho\; \mathbb{I}\otimes\sigma_z\otimes\mathbb{I},
\end{equation}

and $\sigma_z$ is the Pauli matrix

\begin{equation}
\sigma_z= \left( \begin{array}{cc} 1 & 0 \\ 0 & -1\end{array}\right).
\end{equation}

The final channel state obtained through our model, will thus be given by:

\begin{widetext}

\begin{equation}
\label{eq:noisy_state}
\begin{array}{c}
\rho^{noisy}_{AB}=\Tr_{\tau_i}[U_{\lambda/2}(\dfrac{\pi}{4})\;
\mathcal{O}_{calc.}^{noisy}[U_{\lambda/2}(\alpha)\;\mathcal{O}_{calc.}^{noisy}[\rho^{SPDC}_{AB}\otimes\ket{\tau_1}\bra{\tau_1}\;]\;U^{\dagger}_{\lambda/2}(\alpha)]\;
U^{\dagger}_{\lambda/2}(\dfrac{\pi}{4})]=\\\\
\dfrac{1}{4} \left( \begin{array}{cccc}
(1-v)\gamma & 0 &0 & 0\\
0 & (2-\gamma +v(-2+3\gamma)) & -2(1-2\delta)^2 v \gamma & 0\\
0& -2(1-2\delta)^2 v \gamma & (1+v)\gamma &0\\
0&0&0& (2+v(2-3\gamma)-\gamma)
\end{array}\right)
\end{array}
\end{equation}

where we defined $\gamma$ accordingly to: 

\begin{equation}
\label{eq:gamma_def}
\gamma=\dfrac{4 \sin^2 2\alpha}{3-\cos 4 \alpha}.
\end{equation}
and where we regain \eqref{sharedstate}, in case $v=\delta=1$, i.e. absence of noise.\\
Given the state in \eqref{eq:noisy_state}, the nonclassicality witness assumes the value:

\begin{equation}
\label{eq:noisy_witness}
W^{noisy}(\gamma,\theta,v,\delta)=1+v^2-2\gamma v^2-(1-\gamma)(1+v^2)\cos \theta - 2 \gamma (1-2 \delta)^2 v^2 \sin \theta.
\end{equation}
\end{widetext}
Since $\theta$ is still a free parameter, we can minimize the witness over this variable. This minimization is obtained when

\begin{equation}
\label{eq:theta_min_witness}
\theta_{min}(\gamma,v,\delta)=\arctan \dfrac{2(1-2\delta)^2 v^2 \gamma}{(1+v^2)(1-\gamma)}.
\end{equation}

\subsection{Noise Parameters Estimation}
We numerically estimated the noise levels of our experimental apparatus by 
analyzing the relationship between the experimental and theoretical estimation of the nonclassicality witness. For each experimental point we performed a numerical minimization over the free parameter $\theta$, in order to obtain a set of optimal experimental values for the witness, given different values of $\gamma$. We then performed a nonlinear fit of these data given the function $W^{noisy}(\gamma,\theta_{min(\gamma,v,\delta)},v,\delta)$ described in \eqref{eq:noisy_witness} and \eqref{eq:theta_min_witness}, where we considered $v$ and $\delta$ as free parameters. This led to the estimation
\begin{equation}
 v\simeq 0.962 \pm 0.070,\; \delta \simeq 0.872 \pm 0.053. 
\end{equation}

\section{Teleportation witness}

In this section we describe the family of teleportation witnesses that we obtain considering teleportation of the states $\{\psi_x\}_{x=0}^5 = \{(\ket{0}+\ket{1})/\sqrt{2},(\ket{0}-\ket{1})/\sqrt{2},(\ket{0}+ i\ket{1})/\sqrt{2}, (\ket{0}- i\ket{1})/\sqrt{2},\ket{0},\ket{1}\}$ over the share state \eqref{sharedstate}. Moreover we consider that Alice applies partial Bell state measurement, given by the measurement operators $\{M_a^{\rV\rA}\}_{a=0}^1 = \{\ketbra{\psi^-}{\psi^-},\openone-\ketbra{\psi^-}{\psi^-}\}$. The witnesses operators are obtained through the method (based on semidefinite-programing) described in Ref. \cite{PRLDA} and are given in Table~\ref{t:werner states}. They correspond to a family of operators that can detect non-classical teleportation. 

\begin{table*}[ht]
$\psi_x$ \vspace{0.1cm}\\
$a$\hspace{0.1cm}\,
\begin{tabular}{c||c|c|c|c|c|c}
	$F_{a|\psi_x}^\rB\!(\theta)$ & $\frac{\ket{0} + \ket{1}}{\sqrt{2}}$ & $\frac{\ket{0} - \ket{1}}{\sqrt{2}}$ & $\frac{\ket{0} + i\ket{1}}{\sqrt{2}}$ & $\frac{\ket{0} - i\ket{1}}{\sqrt{2}}$ & $\ket{0}$ & $\ket{1}$  \\ \hline\hline
0 & $-2\sin \theta \sigma_x$ & $2\sin \theta \sigma_x$ & $-2\sin \theta \sigma_y$ & $-2\sin \theta \sigma_y$ & $2(1 -\cos \theta)(-\sigma_z + \openone)$ & $2(1 +\cos \theta)(\sigma_z + \openone)$ \\ \hline
1 & 0 & 0 & 0 & 0 & 0 & 0
\end{tabular}
\caption{\label{t:werner states} One-parameter family of teleportation witnesses, for $\theta \in (0,\pi/2]$. These witnesses are useful for demonstrating non-classical teleportation obtained by using the state \eqref{sharedstate}. The input states are the eigenvectors of three Pauli operators and partial Bell state measurement is performed by projecting on a singlet state $\ketbra{\psi^-}{\psi^-}$.}
\end{table*}

For states of the form \eqref{sharedstate}, used to teleport the set of states $\{\psi_x\}$, we find
\begin{equation}
\tr\sum_{a,x}F_{a|\psi_x}^\rB\!(\theta)\sigma_{a|\psi_x}^B=-2\gamma\sin\theta + 2(1-\gamma)(1-\cos\theta),
\end{equation}
which must be negative to demonstrate non-classical teleportation. This is the case whenever
\begin{equation}
 \gamma > \frac{1-\cos\theta}{1-\cos\theta + \sin \theta}.
\end{equation}
For each value of $\gamma$, the largest violation is achieved at $\theta = \theta^*$, where $\tan \theta^* = \gamma/(1-\gamma)$, in which case 
\begin{equation}
\tr\sum_{a,x}F_{a|\psi_x}^\rB\!(\theta^*)\sigma_{a|\psi_x}^B=2(1-\gamma)\left(1-\sqrt{1 + \left(\frac{\gamma}{1-\gamma}\right)^2}\right)
\end{equation}
which is strictly negative except when $\gamma = 0$ (i.e. when the state is the product state $\ket{VV}$). Thus all states of the form \eqref{sharedstate} with $\gamma \neq 0$ demonstrate non-classical teleportation, and this can be witnessed by non-classical teleportation witnesses of the form given in Table \ref{t:werner states}.

\begin{figure}
\includegraphics[width=0.75\columnwidth]{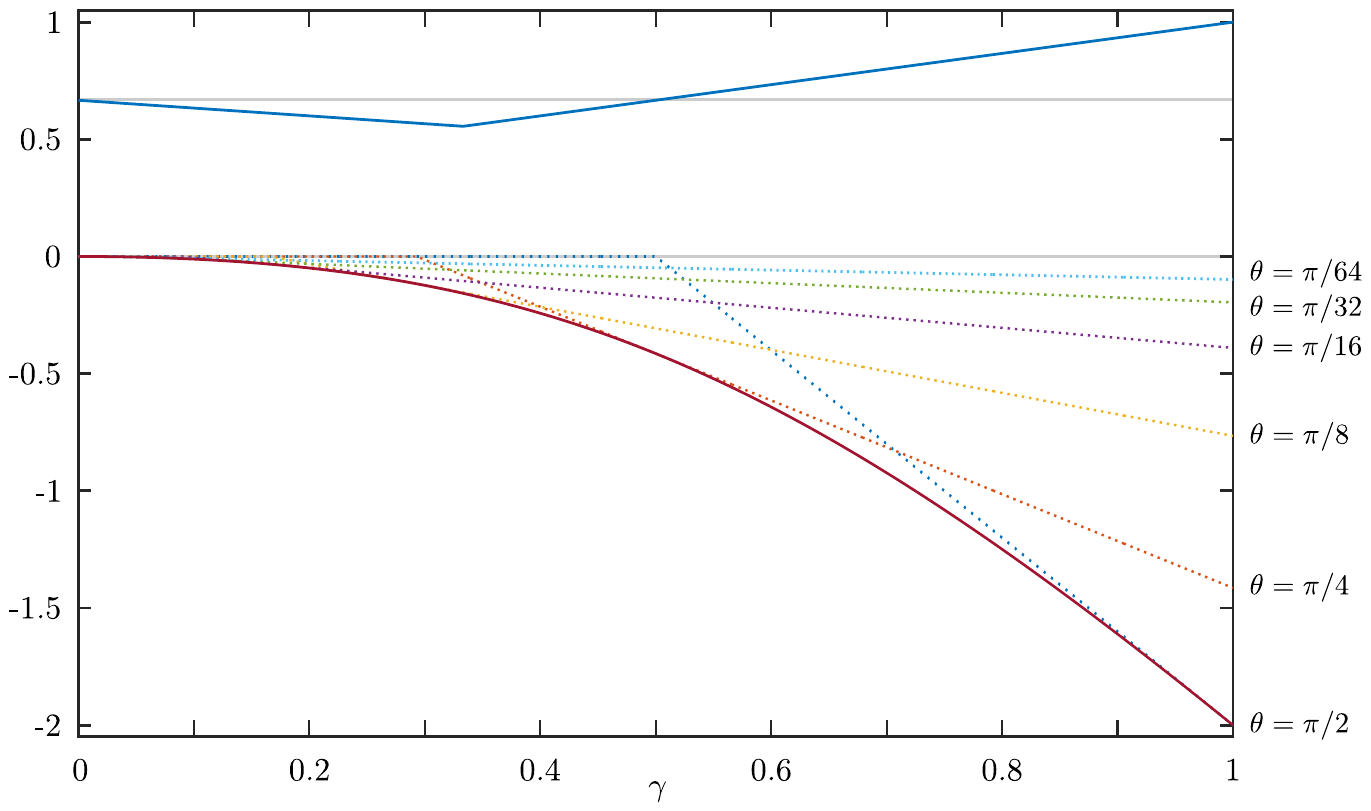}
\caption{The dotted curves show the value obtained for the teleportation witness from Table~\ref{t:werner states}, using the two-qubit state \eqref{sharedstate} and partial Bell state measurement, for various fixed values of $\theta$. The solid curve shows the violation obtained when using the inequality with $\tan \theta^* = \gamma/(1-\gamma)$ as the witness for state with noise $\gamma$, which is the optimal witness to use. The upper solid curve shows the average fidelity of teleportation $\overline{F}_\mathrm{tel}$ as a function of $\gamma$. Classically it is possible to achieve $\overline{F}_\mathrm{cl} = 2/3$ (grey line). Thus, according to this figure of merit, states with $0 \leq \gamma \leq 1/2$ are useless for teleportation.}
\label{fig}
\end{figure}

Notice that the teleportation witness only involves measuring $\sigma_i$ whenever the input state was an eigenvector of $\sigma_i$.  In Fig.~\ref{fig} we plot the violation for various fixed values of $\theta$ (dotted lines), as well as the violation obtained by using the optimal witness ($\theta^*$) for a given $\gamma$ (solid line).

In Fig.~\ref{fig} we also plot the average fidelity of teleportation $\overline{F}_\mathrm{tel}$ for states of the form \eqref{sharedstate}.
Classically it is possible to achieve $\overline{F}_\mathrm{cl} = 2/3$, thus for $0 \leq p \leq 1/2$, states of the form \eqref{sharedstate} cannot beat the classical average fidelity of teleportation, and thus were previously thought to be useless for teleportation. However, as our teleportation witness demonstrates this is not the case, and for all $p > 0$ these states demonstrate non-classical teleportation.